\documentstyle[12pt,psfig]{article}

\newcommand{\apl}{\:^{<}_{\sim}\:}

\begin{document}
\setlength{\baselineskip}{20pt}
\begin{titlepage}
\begin{center}
{\Large \bf MASSIVE NEUTRINO DECAY AND SHAPE OF THE GALACTIC DARK HALO} 

\vskip .1in

{\large Srdjan Samurovi\'c}\footnote{E-mail address: {\it
srdjan@ts.astro.it}}

{\em Dipartimento di Astronomia, \\
Universit\`{a} degli Studi di Trieste, \\
 Via Tiepolo 11, \\
I-34131 Trieste, ITALY}

\vskip .1in

{\large Milan M. \'Cirkovi\'c}\footnote{E-mail address:
{\it cirkovic@mail.ess.sunysb.edu}} 

{\em Astronomical Observatory, \\
Volgina 7, \\
11160 Belgrade-74, YUGOSLAVIA}

{and}

{\em
Department of Physics \& Astronomy\\
S.U.N.Y. at Stony Brook\\
Stony Brook, NY 11794-3800, USA \\ }

\end{center}

\begin{abstract}
In this Letter we investigate the basic assumptions of the decaying dark matter (DDM) theory in the light of recent
advances in observational and theoretical cosmology and physics, 
i.e.~detection of massive astrophysical compact
halo objects (MACHOs) and Super-Kamiokande results. Specifically, the consequences pertaining to the shape of the Milky Way galaxy dark halo
are discussed. We find that, 
by taking into account the values of the main constituent of the mass in DDM theory, massive neutrino, with the mass
 of 30 eV, and lifetime $\sim 10^{23}$ s, the initially proposed value of extreme halo flattening $q\sim 0.2$ is no longer
necessary, and that one can easily accommodate a much larger value of $q\sim 0.6$, that is in accord with all available observational data.
\end{abstract}
\end{titlepage}

\section{Introduction}

It is well known that every spiral galaxy consists of three main parts: central bulge, disk and dark halo (e.g.\ \cite{BM98}). In this Letter we
wish to investigate in more detail the idea that the halo of our Galaxy, Milky Way (in the further text also denoted as the Galaxy) is made mainly of two components: {\it baryonic\/} (MACHOs, see the discussion later) and {\it non-baryonic\/} (massive neutrinos with the mass $m_\nu \sim 30$ eV), and impact of such a composition of matter on the shape of the halo.

The problem of the mass composition in spiral galaxies is well known and still unsolved  (e.g.\ Ref.~\cite{combesetal}). Rotation curve of the Galaxy (Ref.~\cite{mm92}) and non-existence of the Keplerian fall-off
points out that there exists a large amount of matter that
still has to be accounted for. It is a general opinion that a bulk of this
matter resides in the dark halo of each spiral galaxy (Ref.~\cite{BT87,trimble,ashman}), although there are opposite standpoints that suggest the modification of the Newtonian dynamics (MOND) (Refs.~\cite{milgroma,milgromb,milgromc}) that recently produced
 quite good agreements with the observations of 15 rotation curves (Ref.~\cite{deblok}). 
 
Massive neutrinos  remain viable candidates for the "missing light" problem (as  defined by Schramm and Steigman \cite{SS81}). It can be shown that under certain assumptions (Ref.~\cite{TG79}) massive neutrinos with the mass of the order of 30 eV can account for the mass in spiral galaxies (e.g.\ Ref.~\cite{SC96}). These neutrinos play the central role in the
 decaying dark matter (DDM) theory usually associated with the name of Dennis W. Sciama \cite{sciamabook}. 
 
One of the major motivations for undertaking this research is recent
work of the present authors (Ref. \cite{setal}), who marshalled 
lots of evidence pointing to a moderately flattened dark halo $q \sim 0.6$ 
for both our Galaxy and spiral galaxies in general. Most of those 
arguments apply to an essentially baryonic (MACHO or gaseous) halo,
but some are applicable to the general dynamically dominant halo 
component. Sciama's DDM theory is among the plausible and interesting
theories predicting an extreme $q \apl 0.2$ flattening for the dark neutrino component. 

It should be noted that the presence of MACHOs detected through 
microlensing searches does not automatically discard the DDM theory, 
neither its applications in the galactic context. It is still possible 
(albeit more and more difficult to maintain) that MACHOs are dynamically
 insignificant or only marginally significant. The estimates of the
MACHO contribution to the cosmological mass density are still dependent
on several not completely watertight assumptions \cite{fieldsetal,setal}. 
We shall discuss some possibilities for a combined DDM + MACHO model in
the Section 4.

\section{DDM theory: Fundamentals}

According to the DDM theory it is assumed that neutrinos do have masses and that 
the more massive neutrino, $\nu_1$ radiatively decays into
a photon and less massive neutrino, $\nu_2$:
\begin{equation}
\label{decay}
\nu_1\rightarrow \gamma + \nu_2
\end{equation}
(e.g. Ref.~\cite{boehmvogel}). If one further wishes to discuss the kinematics of this equation, one would
obtain the following simple relation \cite{sciamabook}:
\begin{equation}
\label{egamma}
E_\gamma={1\over2}m_\tau\left ( 1-{m_{\nu_{e,\mu}}^2\over m_\tau^2} \right 
),
\end{equation}
where Greek letters in indices are related to the various types of neutrinos, i.e. tau, electron and mu
neutrinos. It is also assumed that 
\begin{equation}
\label{mgor}
m_{\nu_{e,\mu}}\apl 5\; {\rm eV}
\end{equation}
therefore giving the following simple relation:
\begin{equation}
\label{egammasimple}
E_\gamma \sim {1\over 2}m_{\nu_\tau}.
\end{equation}

The basic values of the DDM theory are the following (Ref.~\cite{sciamabook},
 updates in Ref.~\cite{sciama98}):

\begin{itemize}

\item
According to the latest
estimates the mass is: $m_{\nu_\tau}=27.4\pm 0.2$ eV.

\item
 These neutrinos decay radiatively with the lifetime $\tau=2\pm 1
\times 10^{23}$ s. For the latest correction of this value, see the discussion later (based upon 
Ref.~\cite{mohapsci}).

\item
 A decay photon energy is $13.7\pm 0.1$ eV (obtained from the  eq.~[\ref{egammasimple}]),
  so these photons can
ionize hydrogen, but not helium.

\end{itemize}

\section{Flattening in the DDM theory and observational data}
The significant flattening of the halo was introduced by Binney, May and Ostriker (Ref.~\cite{BMO}) who found that it is likely that if the kinematics of the objects in the halo are similar to the kinematics 
of the extreme Population II objects, then the massive halo should 
have the axis ratio $c/a^< _\sim 0.5$. One can now define the 
flattening parameter $\psi$ as $\cos \psi=q=c/a$, i.e.~its cosine determines the shape of the halo $En$. The $En$ notation is related 
to $q$ as $q=1-n/10$.

As for the DDM theory, Sciama \cite{sciama90a} proposed that the 
dark halo had to be extremely flattened $q\sim 0.2-0.3$, in order to 
obtain scale height of electrons responsible for pulsar dispersion 
that is determined by the scale height of the dark matter (i.e.~massive neutrinos). By scale height we
assume the column density on one side of the galactic plane divided by the volume density in the
plane. Later on, Sciama \cite{sciama90b} argued that the scale height of the ionized gas should be reduced to
$900\pm 100$ pc, thus eliminating the need of the strongly flattened halo.
 However, Nordgren, Cordes and Terzian \cite{nordgrenetal} found that the electron scale height is  $670^{+170}_{-140}$ pc, and the mean electron density for the interstellar medium $n_e=0.033\pm 0.002$ ${\rm cm}^{-3}$. This is also in accordance with earlier results of Cordes et al.\ \cite{cordes91}. 
 
This leads us to important questions concerning the ionization problem  in the spiral galaxies (e.g.~Refs.~\cite{sciamabook,SCM98}). We just  mention here that the conventional sources such as O stars
or supernovae have much smaller scale heights ($\sim 100$ pc) than the aforementioned ones of the free electron
component of the interstellar matter. Following Ref.~\cite{sciamabook} we start with the equation that represents
ionization equilibrium:
\begin{equation}
\label{equila}
 {n_\nu (0)\over \tau}=\alpha n_e^2,
\end{equation}
where  $n_\nu (0)$ is the neutrino density near the Sun, $n_e$ is the free electron density, 
$\tau$ is the predicted value of the decaying neutrino lifetime and
$\alpha$ is the hydrogen recombination coefficient excluding transition directly to the ground state
(cf. Ref.~\cite{sciama97}). If one adopts $\alpha=2.6\times 10^{-13}$
${\rm cm}^3 {\rm s}^{-1}$ and $n_e=0.033$ ${\rm cm}^{-3}$
this leads to the following neutrino density:
\begin{equation}
\label{nudens}
n_\nu(0)=2.83\times 10^7\tau_{23}\; {\rm cm}^{-3}
\end{equation}
where $\tau$ is expressed as $\tau=10^{23}\tau_{23}$ s. 
This can give the following value of the mass density $\rho_\nu (0)$, assuming the  neutrino mass of $27.4\pm
0.2$ eV:
\begin{equation}
\label{massdens}
\rho_\nu (0)=1.384\times 10^{-24}\tau _{23}\; {\rm g}\;{\rm cm}^{-3}=0.02\tau _{23}{\cal M}_\odot\; {\rm pc}^{-3}
\end{equation}
where we, as usual, expressed mass density in the Solar masses (${\cal M}_\odot =1.989\times 10^{33}$ g)
 over cubic parsec (1 pc $= 3.0857\times 10^{18}$ cm).

Now we can see why the extremely flattened halo was needed in the DDM theory; namely using equation
(\ref{equila}) one has:
\begin{equation}
\label{nedens}
n_e=\left ({n_\nu \over \alpha \tau}\right )^{1\over 2}
\end{equation}
that gives, after inserting appropriate values, the value for the electron density:
\begin{equation}
\label{nevalue}
0.016\le n_e \le 0.028\; {\rm cm}^{-3}
\end{equation}
where we put Sciama's  value $n_\nu=2\times 10^7$ ${\rm cm}^{-3}$
(cf. Ref.~\cite{sciamabook}) and let lifetime to take the values $\tau=2\pm 1\times 10^{23}$ s.  One can see that there are two ways to achieve agreement of theory with observations: one can either increase $n_\nu$ 
as suggested by  Sciama (Ref.~\cite{sciamabook}), by reducing the
assumed scale height of the neutrino distribution of 8 kpc by a factor of $\sim 4$ to $\sim 2$ kpc which means that one must introduce
extremely flattened halo, or, as we will show one can assume modified form of the equation (\ref{equila}) (see the discussion later).
We first note that we obtained lower value than Sciama (Ref.~\cite{sciama97}) for $\rho_\nu (0)$, 
$0.02\tau _{23}{\cal M}_\odot\; {\rm pc}^{-3}$ vs. $0.03\tau _{23}{\cal M}_\odot\; {\rm pc}^{-3}$, because we used
Nordgren et al. (Ref.~\cite{nordgrenetal}) lower value for $n_e$ (0.033 ${\rm cm}^{-3}$), rather 
 than Sciama 
(0.04 ${\rm cm}^{-3}$). In the Section, we shall see that 
this has profound consequences for flattening, even if we accept 
all other Sciama's premises.

\section{Uncertainties inherent in the flattening parameter 
determination}
We would like now to investigate the general applicability of the decay--ionization equilibrium equation, and in order to do it, we would
like to isolate all possible sources of uncertainty. Therefore we 
propose the following improvements concerning equation (\ref{equila}):
 
1) Additional ionizing sources should be taken into account, regardless on the assumed scale height. These can be O and B stars 
(the question of galactic disks opacity not being completely clear at present), as discussed, for instance, by Mathis \cite{mathis86}
and others \cite{franss}, metagalactic background, cosmic rays, large-scale shocks, etc.

In this case, the equation for decay-ionization equilibrium (\ref{equila}) should read 
\begin{equation}
\label{equilb}
\frac{n_\nu (0)}{\tau} + F = \alpha n_e^2,
\end{equation}
where $F$ is the ionizing contribution of all other sources. Thus, we see that
\begin{equation}
\label{nnu}
n_\nu(0) = (\alpha n_e^2 -F) \tau,
\end{equation}
i.e.\ required neutrino density is {\it decreased}. Accordingly, $\rho_\nu (0)$ is also decreased, and necessity for flattening is decreased, as we
shall see in more detail below. 

2) Assumption of "maximal neutrino" halo is unwarranted, even if desirable. The discovery of
MACHOs, and substantial mass contribution of these objects to the required dynamical mass
of the Galaxy (e.g. Ref.~\cite{fieldsetal}), immediately invalidates this assumption.
So far, using the notation of the Sciama \cite{sciama97}, we have to use the decomposition
\begin{equation}
\label{sigmarota}
\Sigma_{\rm rot} = \Sigma_{\rm rot} (\nu) + \Sigma_{\rm rot} (\rm{other})
\end{equation}
Accordingly, the required lengthscale is decreased by a factor
\begin{equation}
\label{sigmarotb}
\frac{\Sigma_{\rm rot} (\nu)}{\Sigma_{\rm rot}} = f <1.
\end{equation}
This offers an interesting opportunity of building of more complex model
incorporating other observed phenomena. If we suppose that the rest of dynamical mass of the Milky Way halo is made of baryonic MACHOs of
the same population as the one detected in the microlensing searches, 
then one should be able, in principle, to tightly constraint the 
parameter $f$, i.e. the decaying neutrino fraction of the dynamical mass.
This is to be achieved by simultaneous consideration of (i) 
microlensing optical depths and their ratios in various directions,
which would put constraints on the shape of the MACHO component
\cite{setal}; (ii) ionization of Reynolds' layer as in the present
paper; and (iii) constraints on the MACHO abundance stemming from
the Big Bang nucleosynthesis constraints. 

We consider this to be a good bargain, since for price of introducing 
one additional parameter ($f$ ratio) in the theory, we get multiple
advantages: accounting for observed microlensing events and baryonic
dark matter on one hand, and non-baryonic dark matter on the other, as
well as retaining some advantages of the classical DDM theory, like 
explanation of cosmological reionization. Therefore, in the course of future work, we shall try to
show how this extension can be specifically realized.  

3) Column density uncertainty ($\delta \Sigma_{1.1}$ may vary within a factor or $\sim 2$) is introduced. 

4) Galactocentric distance of the Sun, as well as the rotational  velocity
of the LSR are still subject to some uncertainty (e.g.~Ref.\cite{BM98}). 

5) The least significant source of uncertainty is the uncertainty in the mass of decaying neutrino. It is manifested in transition from the 
neutrino number-density (as a discrete value quantifying ionization equilibrium) to the mass-density distribution. 

\begin{figure}
\psfig{file=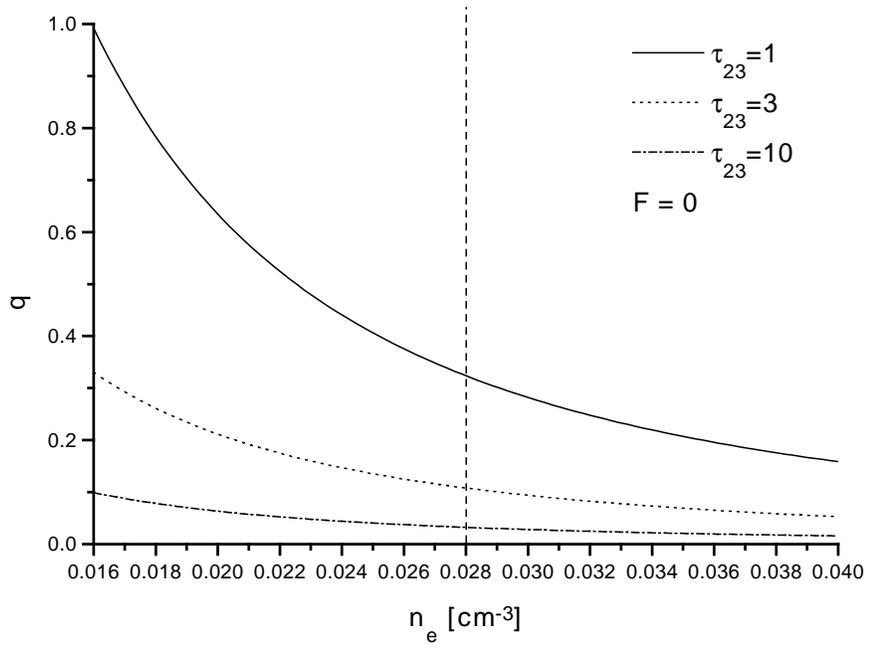,height=13cm}
\caption{The flattening parameter $q$ as a function of the electron density of the Reynolds' layer in the DDM theory. Vertical line denotes the recently obtained observational upper limit on the 
electron density.}
\end{figure}
Now we pass to determination of the flattening parameter of the
neutrino halo, using the method and notation of Sciama \cite{sciama97}.
The vertical lengthscale can be written as
\begin{equation}
\label{sc1}
z_0 = \frac{\Sigma_\nu}{(\alpha n_e^2 -F)\tau m_\nu}.
\end{equation}
Flattening parameter $q$ can be approximated as
\begin{equation}
\label{sc2}
q=\frac{z_0}{r_0},
\end{equation}
where $r_0$ is the lengthscale in the galactic plane. If we accept the
usual form of softened isothermal distribution of neutrino dark matter
\cite{BT87,sciama97}
\begin{equation}
\label{sc3}
\rho_\nu(r) = \frac{\rho_\nu (0) a^2}{a^2 + r^2},
\end{equation}
where $a$ is the core radius. Horizontal lengthscale is, then, given as
\begin{equation}
\label{sc4}
r_0 = \frac{\pi a}{2}.
\end{equation}
Thus, we obtain the general expression for flattening parameter in the
"maximal neutrino" halo assumption with additional ionizing sources:
\begin{equation}
\label{sc5}
q = \frac{2 \Sigma_\nu}{\pi a (\alpha n_e^2 - F) \tau m_\nu}. 
\end{equation}
For $a=8$ kpc, we obtain
\begin{equation}
\label{sc6}
q=6.601 \left\{ \tau_{23} \left[ 2.6 \left( \frac{n_e}{10^{-2} \; {\rm cm}^{-3}}  \right)^2 - \frac{F}{10^{-17}\; {\rm cm}^{-3}\: {\rm s}^{-1}} \right] \right\}^{-1}.
\end{equation} 
Now we have cast the problem in the most tractable form. One may see that the following assumptions 
\begin{enumerate}

\item
purely DDM ionization ($F=0$),

\item
desirable lifetime of the unstable particle ($\tau_{23} \simeq 1$), and

\item
Sciama's original value for the electron density of the Reynolds' layer
($n_e = 0.04$ cm$^{-3}$),

\end{enumerate}
truly lead to extreme flattening of the dark halo ($q = 0.16$). We would
like to investigate the consequences of the relaxing of these assumptions. 
Keeping the assumption (1) and relaxing (2) and (3) leads to results 
shown in the Fig.~1. For the electron density, we use the more 
realistic values of the eq.~({\ref{nevalue}}).

We see that the "classical" value of $\tau_{23}= 1$ {\it does not\/}
warrant an extreme degree of the neutrino halo flattening, contrary
to the claim in the Ref. \cite{sciama97}. Instead, a moderate flattening
$q = 0.4 - 1$ seems to be quite acceptable.

\section{Other ionizing sources}
If we relax the assumption (1), it is necessary to ask: which
ionizing sources are viable for creating the electron density in (\ref{nevalue})? There are several possibilities, main ones being:
\begin{itemize}
\item
O and B stars of the galactic disk;

\item
metagalactic UV ionizing background; 

\item
soft X-ray background, either of galactic or extragalactic origin;

\item
cosmic rays which penetrate the Reynolds' layer;

\item
large-scale ionizing shock waves.
\end{itemize} 
Some of them we can eliminate almost immediately, since we are not interested in the
 detailed ionization structure, only in influence of a specific ionizing source. Cosmic ray ionization, which has played such a prominent role in the first comprehensive ISM model of Field, Goldsmith and Habing in 1969 \cite{fgh}, is now considered to be of secondary importance: an average ionization per H atom is now considered to be $\apl 3 \times 10^{-17}$ s$^{-1}$ (in contradistinction to earlier estimates of $\sim 10^{-15}$ s$^{-1}$). This makes cosmic rays unimportant for the total ionization and heating budget of ISM, with important exception of the interiors of giant 
molecular clouds. If, as we have seen, the density of electrons is
$n_e \sim 10^{-2}$ cm$^{-3}$ (order of magnitude estimates), the total density can not be much higher, certainly $< 10^{-1}$ cm$^{-3}$. 
This makes resulting value of $F$ in the eq.~(\ref{sc6}) quite 
unimportant in comparison with the DDM term, and resulting decrease in flattening is negligible. 

Similar considerations apply to soft X-ray ionization, which was first 
proposed by Silk and Werner \cite{sw69}. The justification is that, like
the cosmic rays, X-rays penetrate vast amount of matter before knocking
out a photoelectron which is subsequently capable of ionizing and
exciting hydrogen. Interestingly enough, for this ionization mechanism,
more electrons are released from He than from H, since helium has larger photoionization cross-section at energies $E_\gamma \sim 100$ eV, situation opposite from the one in case of DDM ionization. However, primary ionization rate for this process is estimated as about $10^{-16}$ 
s$^{-1}$ in an unshielded region \cite{silk}. Thus, resulting value 
of $F$ is $\apl 10^{-17}$ cm$^{-3}$ s$^{-1}$, which does not 
significantly influence the required flattening of the neutrino halo. 
The same conclusion applies {\it a fortiori\/} when we consider effects
of finite opacity of the Reynolds' layer. 
 
What is the most difficult is to estimate the importance of the most 
plausible source of ionization: O and B stars in the Milky
Way disk \cite{ds94}. This uncertainty is reflected in the conclusions of
Mathis \cite{mathis86}, and is connected with the (in)famous 
problem of the opacity of disks of spiral galaxies 
\cite{bm97,letal,xb}, which has not been solved to this day. It seems
clear that the {\it shape\/} of the ionizing spectrum created by 
galactic early-type stars is capable of explaining the ionization
state of the Reynolds' layer, but more detailed modelling of the
global ISM opacity will be necessary to positively discern whether 
this process really takes place. Our conclusion with respect to flattening
seem, however, clear: if a fraction of O and B stars' ionizing flux successfully penetrates dense low-latitude ISM, resulting ionization
would further obviate need for flattening of the hypothetical 
decaying neutrino halo.

\section{Plausible modifications of DDM theory}
Very recently, an attempt has been made to show that the diffuse ionization in the Galaxy can be explained via 
the decaying of sterile neutrinos with a mass of 27.4 eV and lifetime of $\sim 10^{22}$ s \cite{mohapsci}.
Sterile neutrinos appear in the theory as a consequence of the recent Super-Kamiokande result
according to which the atmospheric neutrino anomaly is mainly due to nearly maximal oscillations between $\nu_\mu$ and
$\nu_\mu^s$, where $\nu_\mu^s$ is a sterile neutrino \cite{sciama98a}.
This approach is conceived in order to overcome the perhaps biggest objection to the DDM theory, i.e.~the bulk
of the dark matter is assumed to be in neutrinos. Thus, according to Ref.~\cite{mohapsci}, one can state that
matter contribution to the critical density of the Universe is $\Omega _m\simeq 0.4$, while
the baryonic contribution is $\Omega _B\simeq 0.08$, that leads to halo density of sterile neutrinos of 
$\simeq 2\times 10^6$ ${\rm cm}^{-3}$ that  is 10 times less than the value previously assumed (cf. 
eq.~[\ref{nevalue}]).  According to the 
eq.~[\ref{equila}] this would require that the radiative lifetime is significantly lowered for these sterile
neutrinos, i.e. $\tau \sim 10^{22}$ s. 
However, as we have shown, it is not necessary to consider such radical change of this fundamental 
parameter of the DDM theory -- it is possible to obtain lower concentrations of neutrinos by taking into
account the term $F$, like in eq.~[\ref{equilb}].

\section{Conclusions}
It is clear, from the eq.~(\ref{sc5}), that larger values for $\tau_{23}$,
 require huge amount of flattening, which is
quite unacceptable. For $\tau_{23} \sim 10$, corresponding flattening
parameter is $q<0.1$. This much flattening will have several 
consequences contradictory to the observational data: it is extremely
improbable that such number of microlensing events would have been
observed (Ref.~\cite{melletal}), it would violate the Oort limit (or its absence, see \cite{hipp}), causing either the excessive
local halo density \cite{ggt}, or too small total galactic mass to be consistent with the satellite galaxies measurements \cite{klb,zar1,zar2} (but see \cite{nin90}). Both addition of other ionizing sources and addition of other dynamically important halo components would only act to increase the discrepancy with the minimal tolerable amount of
flattening.\footnote{Strictly speaking, this will apply 
to the addition of another dynamically important halo component 
only if this additional component is not an ionizing source itself, 
which is certainly true for most of the envisaged components (CDM, 
MACHOs, cold gas, etc.). If we stretch our imagination and consider, 
for example, an additional species of decaying particles and call it X, 
then by choosing parameters such that $\Sigma_X /(\tau_X m_X) > 
\Sigma_\nu /(\tau_\nu m_\nu)$ we could increase
the resulting flattening parameter. However, any appeal of the original
DDM (as well as conforming to the Occam's razor) would be lost in this
case, which would pile unwarranted hypotheses and require fine-tuning of
its model parameters.} Thus, we consider this an indication of
non-viability of the large $\tau_{23}$ hypothesis. 

We conclude that in view of all the uncertainties considered, as well as
improved values for several of the observable quantities, flattening of 
the dark halo of the Milky Way can be regarded as neither well-defined nor essential prediction of the DDM theory. 

\vspace{0.7cm}

The authors are happy to express their gratitude to Slobodan Ninkovi\'c and Vesna Milo\v sevi\'c-Zdjelar for useful discussions. SS acknowledges the financial support of the University of Trieste.


\newpage
\begin{thebibliography}{99}
\bibitem{ashman} {K. M. Ashman, {PASP}, {\bf 104}, 1109 (1992).}

\bibitem{BT87} {J. Binney and  S. Tremaine,  {\it 
Galactic Dynamics\/}, (Princeton University Press, Princeton, 1987)}

\bibitem{BMO} {J. Binney, A. May and J. P. Ostriker,   
{MNRAS}, {226}, 149 (1987).}

\bibitem{BM98}{J. Binney and M.  Merrifield, 
 {\it Galactic Astronomy}, (Princeton University Press, Princeton, 1998).}

\bibitem{bm97} {J. Bland-Hawthorn and P. R. Maloney,   PASA, 14, 59 (1997).}

\bibitem{boehmvogel} {F. Boehm and P. Vogel,  
 {\it Physics of Massive Neutrinos}, (Cambridge University Press, Cambridge, 1992).}

\bibitem{combesetal} {F. Combes, P. Boiss\'e, A.  Mazure, A. Blachard, 
  {\it Galaxies and Cosmology\/}, (Sprin\-ger Verlag, New York, 1995).}

\bibitem{cordes91} {J. M. Cordes, J. M. Weisberg, D. A. Frail, S. R. Spangler \& M. Ryan, Nature, 354, 121 (1991).}

\bibitem{deblok} {W. J. G. de Blok and S. S. McGaugh,  preprint
astro-ph/9805120 (1998).}

\bibitem{ds94} {J. B. Dove  and J. M. Shull,  ApJ, 430, 222 (1994).}

\bibitem{fgh} {G. B. Field,  D. W. Goldsmith and H. J. Habing,  ApJ, 155, L149 (1969).}

\bibitem{fieldsetal} {B. D. Fields, K. Freese and D. S.  Graff,  {New Astronomy}, {3}, 347 (1998).}

\bibitem{franss} {C. Fransson and R. A. Chevalier,  ApJ, 296, 35 (1985).}
 
\bibitem{ggt} {E. I. Gates, G. Gyuk and M. S. Turner, ApJ, 449, L123 (1995).}

\bibitem{hipp} {J. Kovalevsky, ARAA, 36, 99 (1998).}

\bibitem{klb} {A. S. Kulessa and D. Lynden-Bell, MNRAS, 255, 105 (1992).}

\bibitem{letal} {C. Leitherer, H. C. Ferguson, T. M.  Heckman and J. D.  Lowenthal, ApJ, 454, L19 (1995).}

\bibitem{mathis86} {J. S. Mathis, ApJ,  {\bf 301}, 423 (1986).}

\bibitem{melletal} {Y. Mellier, F. Bernardeau and  L. van Waerbeke,
to appear in the "IVieme colloque de cosmologie". June 4-6 1997,
 Observatoire de Paris (preprint astro-ph/9802005) (1998).}

\bibitem{mm92} {M. R. Merrifield,   {AJ}, {\bf 103}, 1552 (1992).}

\bibitem{milgroma} {M. Milgrom,  {ApJ}, {\bf 270}, 365 (1983a).}

\bibitem{milgromb} {M. Milgrom, {ApJ}, {\bf 270}, 371 (1983b).}

\bibitem{milgromc} {M. Milgrom,   {ApJ}, {\bf 270}, 384 (1983c).}

\bibitem{mohapsci} {R. N. Mohaparta and D. W. Sciama,  
preprint hep-ph/9811446 (1998).}

\bibitem{nin90} {S. Ninkovi\'c, Bull. Astron. Inst. Czechosl., 41, 236 (1990).}

\bibitem{nordgrenetal} {T. E. Nordgren, J. M. Cordes and Y. Terzian, 
  {AJ}, {\bf 104},} 1465 (1992). 

\bibitem{peeb93} {P. J. E. Peebles,  
 {\it Principles of Physical Cosmology}, (Princeton University Press, Princeton, 1993).}

\bibitem{SC96} {S. Samurovi{\'c} and V. {\v C}elebonovi{\'c}, 
 {Publ. Astron. Obs. Belgrade}, {\bf 54}, 81 (1996).}

\bibitem{SCM98} {S. Samurovi{\'c}, M. M. {\'C}irkovi{\'c}, 
V.   Milo{\v s}evi{\'c}-Zdjelar and
J. Petrovi{\'c},  in 19th Summer School and International Symposium of the Physics of Ionized Gases
(SPIG), August 31  -- September 4,  1998, Zlatibor}, {\it Contributed Papers \& Abstracts of Invited Lectures,
Topical Invited Lectures and Progress reports}, eds. N. Konjevi\'c, M. \'Cuk and I. Videnovi\'c,  635 (1998).

\bibitem{setal} {S. Samurovi\'c, M. M. \'Cirkovi\'c and 
 V. Milo\v sevi\'c-Zdjelar, MNRAS, submitted (1999).}

\bibitem{SS81} {D. N. Schramm and G. Steigman, {ApJ},  {\bf 243}, 1 (1981).}

\bibitem{sciama90a} {D. W. Sciama,   {MNRAS}, {\bf 244}, 1p (1990a).}

\bibitem{sciama90b} {D. W. Sciama, {ApJ}, {\bf 364}, 549 (1990b).}

\bibitem{sciamabook} {D. W. Sciama, 
 {\it Modern Cosmology and the Dark Matter Problem\/}, (Cambridge University Press, Cambridge, 1993).}

\bibitem{sciama95} {D. W. Sciama, in {\it The Physics of the Interstellar Medium and Intergalactic Medium}, ed.~by A. Ferrara et al.~(ASP Conference 
Series, San Francisco, 1995), p. 114}

\bibitem{sciama97} {D. W. Sciama, preprint astro-ph/9704081 (1997).}

\bibitem{sciama98} {D. W. Sciama,   {A \& A}, {\bf 335}, 12 (1998).}

\bibitem{sciama98a} {D. W. Sciama, preprint astro-ph/9811172 (1998).}

\bibitem{silk} {J. Silk, PASP, 85, 207 (1973).}

\bibitem{sw69} {J. Silk and M. Werner, ApJ, 158, 185 (1969).}

\bibitem{TG79} {S. Tremaine and J. E. Gunn,  {Phys. Rev. Lett.}, {\bf 42}, 407 (1979).}

\bibitem{trimble} {V. Trimble,   {ARAA}, {\bf 25}, 425 (1987).}
 
\bibitem{xb} {C. Xu and V. A. Buat,  A \& A, 293, L65 (1995).}

\bibitem{zar1} {D. Zaritsky, R.  Smith, C.  Frenk and  S. D. M. White,   ApJ, 405, 464 (1993)}

\bibitem{zar2} {D. Zaritsky, R.  Smith, C.  Frenk and  S. D. M. White,  ApJ, 478, 39 (1997)}
  
\end{thebibliography}
\end{document}